\documentclass{mem}
\usepackage{natbib}\usepackage{txfonts}\usepackage{balance}
\usepackage{graphicx}
\usepackage[a4paper,breaklinks,dvipdfm]{hyperref}
\idline{75}{282}
\begin{document}
\def\teff{$T\rm_{eff }$}
\def\kms{$\mathrm {km s}^{-1}$}
\def\feh{{\rm [Fe/H]}}

\title{
Ultraviolet properties of Galactic globular clusters with GALEX 
}

   \subtitle{}

\author{
E. \,Dalessandro\inst{1}, R. P. \, Schiavon\inst{2}, F. R. Ferraro\inst{1}, S. T. Sohn\inst{3}, B. Lanzoni\inst{1} \and R. W. O'
Connell\inst{4}
          }

  \offprints{E. Dalessandro}

\institute{
Dipartimento di Fisica e Astronomia, Universit\'a degli Studi di Bologna,
Viale Berti Pichat 6/2, I-40127 Bologna, Italy; \email{emanuele.dalessandr2@unibo.it}
\and
Astrophysics Research Institute, Liverpool John Moores University,
Twelve Quays House, Egerton Wharf, Birkenhead CH41 1LD, UK
\and
Space Telescope Science Institute, 3700 San Martin Dr., Baltimore, MD, 
21218, USA
\and
Astronomy Department, University of Virginia, P.O. Box 400325,
Charlottesville, VA 22904, USA 
}

\authorrunning{Dalessandro }

\titlerunning{UV properties of GGCs}

\abstract{We present ultraviolet (UV) integrated colors of 44 Galactic globular
clusters (GGCs) observed with the Galaxy Evolution Explorer (GALEX) in
both $FUV$ and $NUV$ bands. 
We find for 
the first time that 
GCs associated with the Sagittarius dwarf galaxy have $(FUV-V)$ colors 
systematically redder than GGCs with the same metallicity. 
M31 GCs show almost the same UV colors as GGCs, while M87 are systematically bluer.
We speculate about the presence of an interesting trend,
suggesting that the UV color of GCs may be correlated with 
the mass of the host galaxy, in the sense that more massive 
galaxies possess bluer clusters.

}
\maketitle{}

\section{Introduction}

The main contributors to the UV emission from any stellar system are
the hottest stars. Indeed extreme horizontal branch (HB) and post-HB stars are well
known to be among the hottest stellar populations in globular clusters
(GCs) and contribute substantially to the UV radiation observed from
old stellar systems (Greggio and Renzini 1990;
Dorman et al. 1995 - hereafter DOR95). 
The relative contributions of the various types of stars and
the factors that might lead to larger or smaller populations of
UV-bright stars have remained an open question (Greggio and Renzini 1990;
DOR95; Rich et al. 2005; Sohn et al. 2006). \\
In distant extragalactic systems one can ordinarily observe only the
integrated light of unresolved stellar populations, from which the
hope is to gain knowledge about the underlying stellar population. Galactic
globular clusters (GGCs)
play an important role in understanding the integrated UV colors of
extragalactic systems, especially the so called "UV-upturn" observed in the spectral energy 
distributions of elliptical galaxies (Code \& Welch 1979).
In fact,
GCs are the closest example in nature to a single stellar
population (SSP), moreover they span a large range of metallicities,
a small range of ages, and perhaps some range of helium
abundance. 
GGCs are the ideal target to study the impact of hot and bright populations (as
the AGB-manqu\'e stars) on the integrated UV light and 
represent a crucial local template for comparison with integrated
properties of distant extragalactic systems. \\
With this motivation in mind we observed 44 GGCs with the \emph{Galaxy Evolution Explorer} (GALEX)
in $FUV$ and $NUV$.
This is the largest homogeneous sample ever collected for GGCs in
UV so far. 
We obtained resolved photometry in Schiavon et al. (2012) and integrated colors 
in Dalessandro et al. (2012).

\section{The UV integrated colors.}

The integrated UV magnitudes have been obtained by fitting the observed 
surface brightness profiles in each available band with proper King models 
(see Dalessandro et al. 2012 for details).
In order to investigate any possible link between the UV colors and
chemical compositions of the Milky Way GCs, we adopted the
\feh\ values quoted by Carretta et al. (2009).

\begin{figure}[!]
\vspace{0.2cm}
\begin{center}
\resizebox{4.5 cm}{!}{\includegraphics[clip=true]{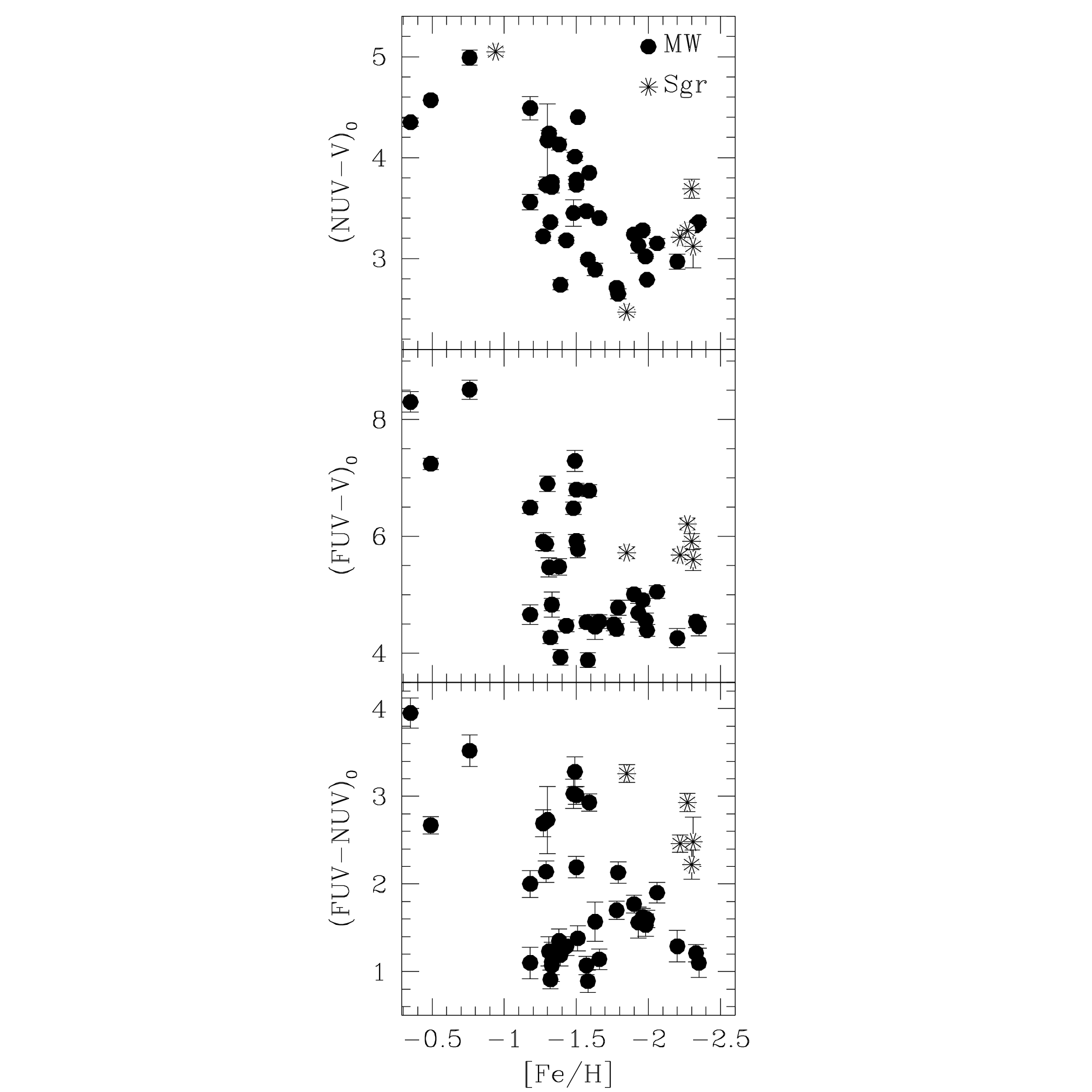}}
\caption{\footnotesize UV integrated colors as a function of metallicity in Carretta et al. (2009)
  scale. Clusters possibly connected with the Sagittarius stream are
  plotted as asterisks.
}
\label{fig1}
\end{center}
\end{figure}

From the top panel of  Fig.~1 one can see that,
for the GCs in our sample, $(NUV-V)_0$ decreases by about 2
magnitudes as \feh\ decreases from $\sim$ -0.7 to $\sim$ --1.5. For
smaller values of \feh, $(NUV-V)_0$ is roughly constant, or perhaps
increases slightly as \feh\ decreases.  This clear, although
non-monotonic, trend of $(NUV-V)_0$ with metallicity is confirmed by a
Spearman correlation rank test, according to which the probability of
a correlation is 99.99\% ($>$4$\sigma$).
Interpretation of the dependence of colors involving $FUV$ magnitudes
as a function of \feh\ requires a little more care.
$(FUV-V)_0$ varies by almost 5 magnitudes.
Indeed, a Spearman test gives a probability
P$\sim$99$\%$ (corresponding to $\sim$2.3 -2.5$\sigma$) that $(FUV-V)_0$
is correlated with metallicity.  The Kendall test gives correlation
probability of $\sim$2.1$\sigma$. At
face value, therefore, $(FUV-V)_0$ correlates with \feh\ in a
similar way as $(NUV-V)_0$, although in a less strong or noisier
fashion.

In summary, then, one finds that the behavior of GGCs in the
$(FUV-V)_0$--\feh\ and $(NUV-V)_0$--\feh\ planes is essentially the
same.  On both planes, three sub-families of clusters can be
recognized: 1) GGCs with $\feh$$>$-1.0, which are predominantly red;
2) GGCs with  -1.5$<$ $\feh$$<$-1.0, the
``second parameter region'',
 where GGCs have a wide range of colors,
about $\sim$2 mag in $(NUV-V)_0$ and $\sim$4 mag in $(FUV-V)_0$; and
3) GGCs with $\feh$$<$ -1.5, which are all blue.  It is worth
noticing that, intermediate-metallicity ([Fe/H]$\approx$-1.5) clusters
are the bluest in the three colors combinations. This was also
highlighted by DOR95. The extension of their HBs (see Schiavon et al. 2012) is
compatible with their integrated colors. On average, the metal-poor
($\feh$$<$-1.7) GCs have redder HBs than the intermediate ones.

\subsection{GCs in the Sagittarius Stream} 

Careful inspection of the $(FUV-NUV)_0$ or $(FUV-V)_0$ \emph{vs}
\feh\ plots in Fig.~1 clearly reveals that the color spread
at $\feh$$<$-1.5 is due to a subset of clusters (plotted as asterisks),
which are systematically redder by $\sim$1.5 and 1.0 mag
in $(FUV-NUV)_0$ and $(FUV-V)_0$, respectively, than the other GCs in
the same metallicity regime.
Interestingly these clusters (NGC~4590, NGC~5053,
NGC~5466, Arp~2 and Terzan~8) are potentially connected with the
Sagittarius dwarf galaxy stream (Law \& Majewski 2010), and thus may
have an extra-Galactic origin. The other candidate Sagittarius GC is the
relatively metal-rich ([Fe/H]=$-0.94$) Palomar~12, for which we were not able
to get $FUV$ magnitude.
These clusters are coeval within the uncertainties (Salaris \& Weiss 2002 and
Dotter et al. 2010).\\
We used the {\it R'-parameter}
reported by Gratton et al. (2010) to highlight possible differences.
The {\it R'-parameter} is defined as the ratio
between the number of HB stars and that of RGBs brighter than the HB level magnitude
$V_{\rm HB}+1$.
This quantity is an indirect estimate of Helium abundances ($Y$).
It is interesting to note that these clusters have {\it R'} values
smaller than other clusters with similar metallicity [Fe/H]$<$-1.5.
Given the statistical uncertainties of these measurements, the difference in {\it R'} between
any two given clusters is somewhat uncertain.
Therefore we performed a t-test to check the significance of the difference between
the mean values of the
two distributions. We find that for the clusters potentially connected with the Sagittarius 
stream $<$R'$>$=0.48$\pm0.01$ while for GGCs $<$R'$>$=0.74$\pm0.18$. 
The t-test gives a probability P$>99.9\%$ that they are different.
This difference might be an indication that those 
clusters have lower He abundances than GGCs in the same metallicity regime, and this is likely the 
main responsible of the differences in $FUV$ integrated colors.

\begin{figure}[!]
\vspace{0.2cm}
\begin{center}
\includegraphics[height=10.0cm,width=6.5cm]{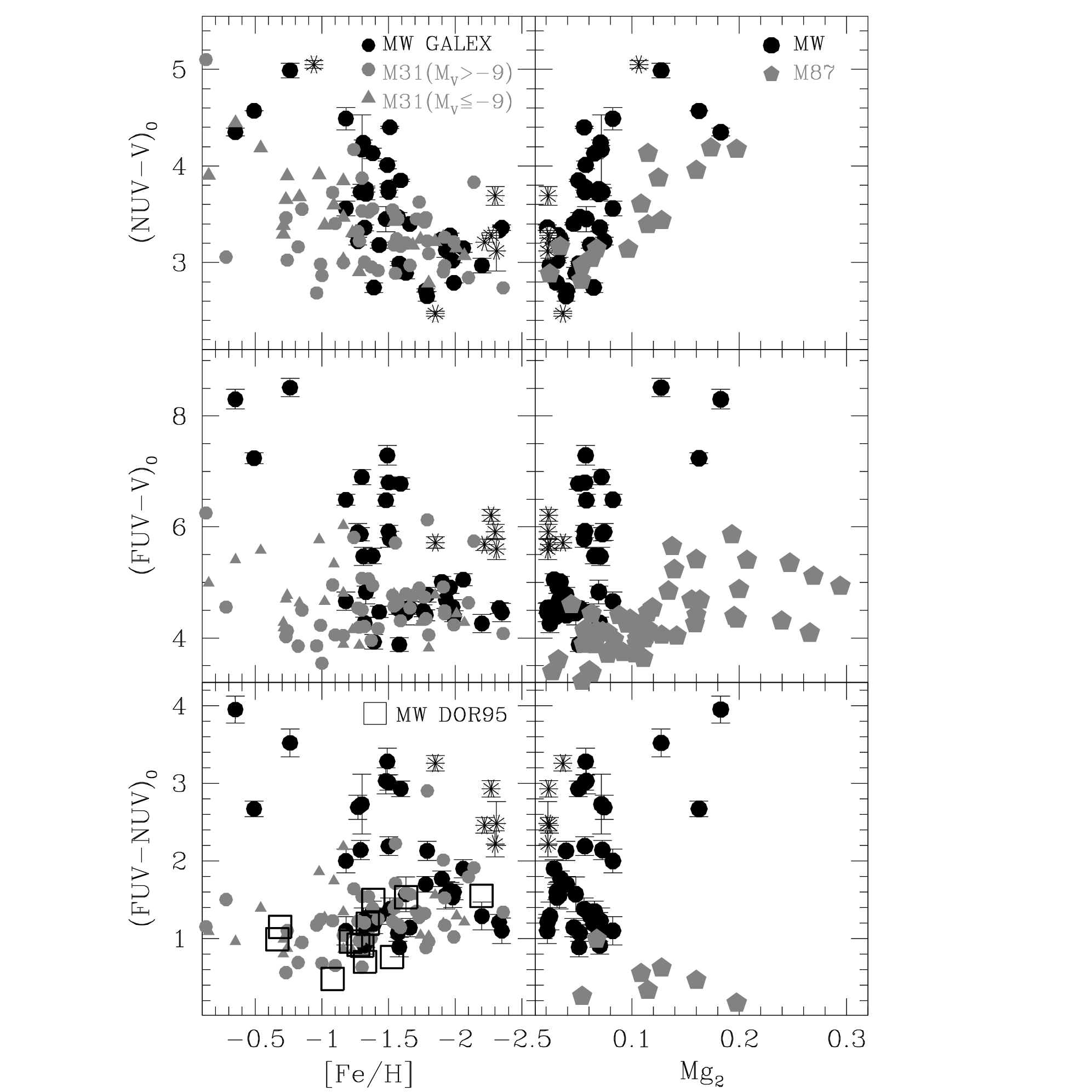}
\caption{\footnotesize {\it Left panel}. GALEX colors of our GGC sample (black) compared to the GCs in
  M31 (from Kang et al. 2011). In the lower panel our GGC sample has been
  supplemented with clusters observed with ANS and OAO2 by DOR95 (open
  squares). {\it Right panel}. UV colors of GGCs (black dots and asterisks) compared to those of M87
(grey pentagons) as a function of the $Mg_2$ metallicity index (Sohn et al. 2006).}
\label{fig2}
\end{center}
\end{figure}

\subsection{Comparison with GCs in M31 and M87}

We compared UV colors of old M31 GCs obtained with GALEX by Kang et al (2012)
with those of GGCs (see left panel of Fig.~2).
M31 seems to show a lack of red clusters with respect to the Galaxy.
However, this is likely due to the limited sensitivity of GALEX to detect
relatively red populations in distant systems.
Hence for the comparison we focus on the bluest systems,
$(FUV-NUV)_0$$<$1.5, $(FUV-V)_0$$<$5 and $(NUV-V)_0$$<$3.5,
which, for the GGCs, correspond to a metallicity range -2.5$<$$\feh$
$<$-1.0. In this metallicity regime the distributions in the Milky Way and
in M31 are quite similar.  The bluest colors reached
are essentially the same, and the distributions show little variations
with metallicity. The case is very different at higher metallicity, $\feh$$>$-1. 
In order to
make the comparison with M31 GCs as complete as possible, we have
supplemented our GALEX sample with 12 additional GGCs from DOR95 not observed by GALEX. 
In the Milky Way
sample there are only red GGCs, while in M31 there are many blue
GCs. In the left panel of Fig.~2 it is possible to note that roughly half of the blue,
metal-rich M31 GCs are
indeed quite massive ($M_V$$\leq$-9).  Hence, the relative paucity of
hot, metal-rich GCs in the Milky Way could be due in part (but only in part) to
the fact that there are only two massive metal-rich clusters in our
supplemented sample. It is also possible that many GGCs with high metallicity 
and a blue HB are missed because of their location towards highly extinguished 
regions of the Galaxy.\\
We also compare GGCs colors with those obtained for
the giant elliptical galaxy M87 using {\it Hubble Space Telescope} 
STIS images (Sohn et al. 2006). As appears in the right panel
of Fig.~2, M87 GCs are on average
bluer by $\sim$1.5 mag both in $(FUV-NUV)_0$ and $(FUV-V)_0$, while
they do not show any appreciable difference in $(NUV-V)_0$.

\section{Conclusions}

From the comparison between GGCs and those belonging to three other
galaxies (the Sagittarius dwarf, M31 and M87), different behaviors
emerged. In fact the clusters associated with the Sagittarius dwarf are
 on average redder than
the MW ones, the M31 clusters have colors comparable to
those of the GGCs, while the M87 star systems are bluer. 
We note that there may be a possible trend between the mass 
of the host galaxy and
the color distribution of its globulars, in the sense that the higher is
the galaxy mass, the bluer are the GC UV colors.
We argued that most
of the observed differences between colors involving the $FUV$ band
are explainable invoking different Helium contents.  This would lead
us to speculatively think that galaxies with larger masses may have, on
average, more He-rich populations. In that case, He abundance differences could be 
a by-product of chemical evolution differences, in some way connected 
to the mass of the host galaxy. This could be also connected with the 
formation and dynamical history of
clusters in galaxies with different masses, as suggested by Valcarce \& Catelan (2011). 
In particular they argue
that clusters hosted by more massive galaxies are more likely to undergo a more complex history of
star formation thus having a larger spread in stellar populations properties.

\begin{acknowledgements}
This research is part of the project COSMIC-LAB funded by the
European Research Council (under contract ERC-2010-AdG-267675).
Photometric catalogs and integrated colors can be downloaded from http://www.cosmic-lab.eu.
\end{acknowledgements}

\bibliographystyle{aa}

\begin{thebibliography}{}


\bibitem[Carretta et al.(2009)]{2009A&A...508..695C} Carretta, E., Bragaglia, A., Gratton, R., D'Orazi, V., \& Lucatello, S.\ 2009, \aap, 508, 695 




\bibitem[Code 
\& Welch(1979)]{1979ApJ...228...95C} Code, A.~D., \& Welch, G.~A.\ 1979, \apj, 228, 95 

\bibitem[Dalessandro et al.(2012)]{2012AJ....144..126D} Dalessandro, E., 
Schiavon, R.~P., Rood, R.~T., et al.\ 2012, \aj, 144, 126 

\bibitem[Dorman et al.(1995)]{1995ApJ...442..105D} Dorman, B., O'Connell, 
R.~W., \& Rood, R.~T.\ 1995, \apj, 442, 105 (DOR95)


\bibitem[Dotter et al.(2010)]{dotter10} Dotter, A., et al.\ 
2010, \apj, 708, 698 

\bibitem[Gratton et 
al.(2010)]{gratton10} Gratton, R.~G., Carretta, E., Bragaglia, A., Lucatello, S., \& D'Orazi, V.\ 2010, \aap, 517, A81 

\bibitem[Greggio 
\& Renzini(1990)]{1990ApJ...364...35G} Greggio, L., \& Renzini, A.\ 1990, \apj, 364, 35 

\bibitem[Kang et al.(2012)]{2012ApJS..199...37K} Kang, Y., Rey, S.-C., 
Bianchi, L., et al.\ 2012, \apjs, 199, 37 

\bibitem[Law 
\& Majewski(2010)]{sgrmass} Law, D.~R., \& Majewski, S.~R.\ 2010, \apj, 718, 1128


\bibitem[Rich et al.(2005)]{2005ApJ...619L.107R} Rich, R.~M., et al.\ 2005, 
\apjl, 619, L107

\bibitem[Salaris 
\& Weiss(2002)]{2002A&A...388..492S} Salaris, M., \& Weiss, A.\ 2002, \aap, 388, 492 


\bibitem[Schiavon et al.(2012)]{2012AJ....143..121S} Schiavon, R.~P., 
Dalessandro, E., Sohn, S.~T., et al.\ 2012, \aj, 143, 121 

\bibitem[Sohn et al.(2006)]{sohnm87} Sohn, S.~T., O'Connell, 
R.~W., Kundu, A., Landsman, W.~B., Burstein, D., Bohlin, R.~C., Frogel, 
J.~A., \& Rose, J.~A.\ 2006, \aj, 131, 866

\bibitem[Valcarce 
\& Catelan(2011)]{2011A&A...533A.120V} Valcarce, A.~A.~R., \& Catelan, M.\ 2011, \aap, 533, A120 


\end{thebibliography}

\end{document}